\documentclass[lettersize,journal]{IEEEtran}
\usepackage{cite}
\usepackage{amsmath,amssymb,amsfonts}
\usepackage{algorithmic}
\usepackage{algorithm}
\usepackage{graphicx}
\usepackage{textcomp}
\usepackage{xcolor}
\newcommand{\RNum}[1]{\uppercase\expandafter{\romannumeral #1\relax}}
\usepackage{subfigure}
\usepackage{tabularx}
\usepackage{booktabs}
\usepackage{array}
\usepackage{stfloats}
\usepackage{url}
\usepackage{verbatim}
\usepackage{graphicx}
\usepackage{cite}
\begin{document}

\newtheorem{theorem}{Theorem}[section]
\newtheorem{corollary}{Corollary}[theorem]    
\newtheorem{lemma}[theorem]{Lemma}

\title{Distributed Quantized Detection of Sparse Signals Under Byzantine Attacks }

\author{Chen~Quan, Yunghsiang~S.~Han,~\IEEEmembership{Fellow,~IEEE}, Baocheng~Geng and~Pramod~K.~Varshney,~\IEEEmembership{Life~Fellow,~IEEE}
\thanks{C. Quan and P. K. Varshney are with the Department of Electrical Engineering and Computer
Science, Syracuse University, Syracuse, NY 13244 USA (e-mail: \{chquan,varshney\}@syr.edu).}
\thanks{Y. S. Han is with the Shenzhen Institute for Advanced Study, University of Electronic Science and Technology of China, Shenzhen, China (e-mail: yunghsiangh@gmail.com).}
\thanks{B. Geng is with the Department of Computer Science, University of Alabama at Birmingham, Birmingham, AL 35294 USA  (e-mail: bgeng@uab.edu).}
}
\maketitle

\IEEEpeerreviewmaketitle
\begin{abstract}
This paper investigates distributed detection of sparse stochastic signals with quantized measurements under Byzantine attacks. Under this type of attack, sensors in the networks might send falsified data to degrade system performance. The Bernoulli-Gaussian (BG) distribution in terms of the sparsity degree of the stochastic signal is utilized for modeling the sparsity of signals. Several detectors with improved detection performance are proposed by incorporating the estimated attack parameters into the detection process. First, we propose the generalized likelihood ratio test with reference sensors (GLRTRS) and the locally most powerful test with reference sensors (LMPTRS) detectors with adaptive thresholds, given that the sparsity degree and the attack parameters are unknown. Our simulation results show that the LMPTRS and GLRTRS detectors outperform the LMPT and GLRT detectors proposed for an attack-free environment and are more robust against attacks. The proposed detectors can achieve the detection performance close to the benchmark likelihood ratio test (LRT) detector, which has perfect knowledge of the attack parameters and sparsity degree. When the fraction of Byzantine nodes are assumed to be known, we can further improve the system's detection performance. We propose the enhanced LMPTRS (E-LMPTRS)
and enhanced GLRTRS (E-GLRTRS) detectors by filtering out potential malicious
sensors with the knowledge of the fraction of Byzantine nodes in the network. Simulation results show the superiority of proposed enhanced detectors over LMPTRS and GLRTRS detectors.
\end{abstract}
\section{Introduction}
With the development of compressive sensing (CS) \cite{fornasier2015compressive,zhang2018kind,donoho2006compressed} in recent years, the sensors in sensor networks often send low-dimensional compressed measurements to the Fusion Center (FC) instead of high-dimensional sparse data, thereby improving bandwidth efficiency and reducing the communication overhead. A high-dimensional signal is sparse when only a few entries in the signal are non-zeros, and others are zeros. Under the CS framework, the reconstruction and the detection of sparse signals have received considerable attention. In this paper, we are interested in detecting compressed sparse signals. 

The problem of compressed sparse signal detection in sensor networks has been studied in the literature\cite{duarte2006sparse,wimalajeewa2016sparse,zayyani2016double,hariri2017compressive,wang2018detection,wang2019distributed,li2019distributed,mohammadi2022generalized,wang2019detection}. In these studies, the recovery of sparse signals was not necessarily required. In \cite{duarte2006sparse,wimalajeewa2016sparse,zayyani2016double}, partly or completely reconstructed sparse signals are required to derive the test statistics for sparse signal detection, while in \cite{hariri2017compressive,wang2018detection,wang2019distributed,li2019distributed,mohammadi2022generalized}, the test statistics are directly derived from compressed measurements to perform sparse signal detection. In \cite{duarte2006sparse} and \cite{wimalajeewa2016sparse}, the authors proposed orthogonal matching pursuit (OMP) algorithms to detect the presence of a sparse signal based on single measurement vectors (SMVs) and multiple measurement vectors (MMVs), respectively, by estimating only a fraction of the support set of a
sparse signal. In \cite{hariri2017compressive}, the Bernoulli-Gaussian (BG) distribution was utilized to model the random sparsity of sparse signals, and the generalized likelihood ratio test (GLRT) was proposed to address the unknown degree of sparsity. Note that under the BG model (which is widely used to model the sparsity of signals \cite{hariri2017compressive,korki2016iterative,soussen2011bernoulli}), the sparse signal has a zero sparsity degree if the signal is absent, but a nonzero sparsity degree that approaches zero if the signal is present. Since the sparsity degree is nonnegative and close to zero, parameter testing based on the sparsity degree can be employed for sparse signal detection by formulating the problem as a one-sided and close hypothesis testing problem. It is worth emphasizing that, although the GLRT strategy is commonly used in the context of signal detection \cite{fang2013one,gao2014quantizer,besson2016generalized}, it lacks solid optimality properties, and its performance may degrade in the case of close hypothesis testing \cite{kay2002optimal}. In \cite{wang2018detection}, instead of GLRT, a method based on the locally most powerful test (LMPT), which is a popular tool for the problems of one-sided and close hypothesis testing, was proposed for detecting sparse signals in sensor networks. The test statistic of the LMPT detector was directly derived from the compressed measurements without any signal recovery. The detectors proposed in \cite{duarte2006sparse,wimalajeewa2016sparse,hariri2017compressive,wang2018detection} assume that the raw signals are transmitted within the network. The limited bandwidth constraints in practice, however, necessitate consideration of the case where only quantized data is transmitted over sensor networks. Next, we discuss sparse signal detectors that are based on quantized data.

A two-stage detector based on the generalized likelihood ratio test (GLRT), where sparse signal recovery is integrated into the detection framework, is proposed in \cite{zayyani2016double} for sparse signal detection from 1-bit CS-based measurements. However, due to substantial information loss caused by 1-bit quantization, there exists a considerable performance gap between the detector based on 1-bit measurements and the clairvoyant detector based on analog measurements \cite{gao2014quantizer}. To mitigate this problem, the authors in \cite{wang2019detection} proposed a quantized LMPT detector that enables the system to achieve detection performance comparable to a clairvoyant LMPT detector by selecting a reasonable number of reference sensors. An extension of the above design with generalized Gaussian noise is presented in \cite{wang2019distributed}. In \cite{li2019distributed}, an improved-1-bit LMPT detector is proposed that optimizes the quantization procedure, and a reduced number of sensor nodes is required to compensate for the loss of performance caused by 1-bit quantization. The authors of \cite{mohammadi2022generalized} proposed a generalized LMPT detector for distributed detection of a phenomenon of interest (PoI) whose position and emitted power are unknown.

In this paper, unlike the previously proposed LMPT detectors \cite{wang2018detection,mohammadi2022generalized,li2019distributed,wang2019distributed}, and the commonly used GLRT detector \cite{hariri2017compressive,zayyani2016double} for sparse signal detection proposed under the assumption of an attack-free environment, we consider the robustness of these detectors as well as their detection performance in the presence of Byzantine attacks. When the system is under Byzantine attacks, one or more sensors in the network may get compromised and may send falsified data to the FC to degrade the detection performance of the system \cite{vempaty2018secure,quan2022enhanced,lin2020minimum,wu2018generalized,liu2018secure,fu2019entropy}.  More specifically, we consider the generalized GLRT detector and the previously proposed LMPT detectors with unknown random sparse signals operating under Byzantine attacks. The random unknown sparse signals are still characterized by the BG distribution as in \cite{wang2018detection,mohammadi2022generalized,li2019distributed,wang2019distributed,hariri2017compressive,korki2016iterative,soussen2011bernoulli,zayyani2016double}. We evaluate the performance of the generalized GLRT detector and the LMPT detectors when they operate under Byzantine attacks. The simulation results show that the detectors are vulnerable to Byzantine attacks because their performance degrades. 
Intuitively, we need more information about the attack parameters to improve the robustness of the previously mentioned detectors in the presence of attacks. In order to attain this goal, we develop a framework for estimating unknown parameters that are inspired by the works in \cite{sartzetakis2019accurate,sartzetakis2017improving}, where supervised machine learning (ML) was utilized as quality of transmission (QoT) estimator for optical transport networks. 
In \cite{sartzetakis2019accurate} and \cite{sartzetakis2017improving}, a part of the total data is used to obtain a sufficiently accurate estimate of the unknown underlying parameters. 

In this work, a subset of the sensors is randomly selected, with their decisions serving as training samples for estimating the unknown attack parameters in the network. We introduce the notion of reference sensors to represent those sensors whose local decisions help as training samples in our problem and propose the generalized likelihood ratio test with reference sensors (GLRTRS) and the locally most powerful test with reference sensors (LMPTRS) with adaptive thresholds, given that the sparsity degree and the attack parameter are unknown. The proposed detectors allow us to yield excellent system performance without knowing the attack parameters.
We assume the Byzantines do not have perfect knowledge about the actual state of the phenomenon of interest and attack based on their local decisions. The simulation results show that the LMPTRS and the GLRTRS detectors outperform the LMPT and the GLRT detectors under attack and can achieve the detection performance close to the benchmark likelihood ratios test (LRT) detector, which has perfect knowledge of all the information, i.e., the attack parameters and sparsity degree. When the fraction of Byzantines in the networks is assumed to be known, enhanced LMPTRS (E-LMPTRS) and enhanced GLRTRS (E-GLRTRS) detectors are proposed to further improve the detection performance of the system by filtering out potential malicious sensors. Simulation results show that the proposed enhanced detectors outperform LMPTRS and GLRTRS detectors.

The paper is organized as follows. We present our system model in Section \ref{sec:system_model}. We evaluate the performance of GLRT and quantized LMPT detectors under Byzantine attacks in Section \ref{sec:proposed}. The robust GLRTRS, LMPTRS, E-GLRTRS, and E-LMPTRS detectors with adaptive thresholds are proposed in Section \ref{sec:proposed2}. We present our simulation results in Section \ref{sec:simulation} and conclude in Section \ref{sec:conclusion}.

\section{System model}\label{sec:system_model}
Consider a binary hypothesis testing problem of detecting sparse signals where hypotheses $\mathcal{H}_1$ and $\mathcal{H}_0$ indicate the presence and absence of the sparse signal, respectively. We consider a distributed network consisting of one fusion center (FC) and N sensors that observe the signals that share the joint sparsity pattern. Let $y_i$ be the received observation at sensor $i\in\{1,2,\dots,N\}$. We assume that all the observations are independent and identically distributed (i.i.d) conditioned on the hypotheses. For sensor $i$, the observation $y_i$ is modeled as 
\begin{align}
y_i = 	\begin{cases}
		n_i&\text{under $\mathcal{H}_0$}\\
		\mathbf{h_i}^{T}\mathbf{x_i}+n_i&\text{under $\mathcal{H}_1$},    	
		\end{cases}
\end{align}
where $\mathbf{x_i}\in\Re^{M\times1}$ is the sparse signal  received by sensor $i$, $\mathbf{h_i}\in\Re^{M\times1}$ is the random linear operator of sensor $i$, $n_i$ is Gaussian noise with zero mean and variance $\sigma_n^2$ and $(\cdot)^T$ denotes the transpose operation. Based on the received compressed measurements $\{y_i\}_{i=1}^N$ from all the sensors, the FC makes a global decision about the absence or presence of the sparse signals.

We adopt the BG distribution introduced in \cite{hariri2017compressive,korki2016iterative,soussen2011bernoulli} to model the sparse signals where the joint sparsity pattern is shared among all the signals observed by the sensors. The locations of nonzero coefficients in $x_i$ are assumed to be the same across all the sensors. Let $\mathbf{s}\in\Re^{M\times1}$ describe the joint sparsity pattern of $\{\mathbf{x}_i\}_{i=1}^N$, where
\begin{align}
    \begin{cases}
		s_m=1,&\text{for $\{x_{i,m}\neq 0, i=1,2,\dots,N\}$}\\
		s_m=0,&\text{for $\{x_{i,m}=0, i=1,2,\dots,N\}$}
	\end{cases}
\end{align}
for $m=1,2,\dots,M$. $\{s_m\}_{m=1}^M$ are assumed to be i.i.d. Bernoulli random variables with a common parameter $p$ ($p\rightarrow0^{+}$), where $P(s_m=1)=p$ and $P(s_m=0)=1-p$. In other words, $p$ represents the sparsity degree of the sparse signal $\mathbf{x}_i$ for $\forall i\in\{1,2,\dots,N\}$. Moreover, each element of $\mathbf{x}_i$ is assumed to follow an i.i.d. Gaussian distribution $\mathcal{N}(0,\sigma^2_x)$\cite{liang2008sensing}. Therefore, the BG distribution is imposed on $x_{i,m}$ as
\begin{align}
    x_{i,m}\sim p\mathcal{N}(0,\sigma^2_x)+(1-p)\delta(x_{i,m}),
\end{align}
where $\delta(\cdot)$ is the Dirac delta function. Due to the limited bandwidth, the sensors send the quantized observations instead of raw observations $\{y_i\}_{i=1}^N$ to the FC. We assume that a fraction $\alpha$ of the total $N$ sensors, namely, $\alpha N$ sensors, are compromised by the Byzantines. We also assume that the compromised sensors are uniformly distributed in the network. In other words, a sensor $i$ can be honest (H) with probability $1-\alpha$ or Byzantine (B) with probability $\alpha$. The Byzantines may intentionally send falsified local decisions to the FC with an attack probability, i.e., the probability that Byzantines flip their decision. The fraction of Byzantines $\alpha$ and the probability that Byzantines flip their decision, $P_A$, are considered attack parameters.  Let $\mathbf{z_i}$ denote the actual 
quantized observation at sensor $i\in\{1,2,\dots,N\}$. The $q$-bit quantizer 
at the $i^{th}$ sensor is defined as
\begin{align}\label{eq:quantize}
\mathbf{z_i} = 	\begin{cases}
		\mathbf{v_1}&\tau_{i,0}\leq y_i\leq \tau_{i,1}\\
		\mathbf{v_2}&\tau_{i,1}\leq y_i\leq \tau_{i,2}\\
		\vdots&\vdots\\
		\mathbf{v_{2^q}}&\tau_{i,2^q-1}\leq y_i\leq\tau_{i,2^q}, 	
		\end{cases}
\end{align}
where $\mathbf{v_k}$ is the binary code word with $\mathbf{v_k}\in\{0, 1\}^q$ and $\{\tau_{i,l},l=0,1,2,\dots,2^q\}$ are the quantization thresholds. For example, given $q=2$, we have $\mathbf{v_1}=00$, $\mathbf{v_2}=01$, $\mathbf{v_3}=10$ and $\mathbf{v_4}=11$. Let $\mathbf{u_i}$ be the decision sent to the FC.
If sensor $i$ is honest, we have $P(\mathbf{u_i}=\mathbf{z_i}|i=H)=1$, otherwise we have $P(\mathbf{u_i}\neq\mathbf{z_i}|i=B)=P_A$. Here, the probability density function (PDF) of the local decision $\mathbf{u_i}$ if $i$ is honest is given as 
\begin{align}\label{eq:prob_u_i}
    P(\mathbf{u_i}|i=H,\mathcal{H}_h)=&P(\mathbf{z_i}|i=H,\mathcal{H}_h)\notag\\
    =&\prod_{j=1}^{2^q}P(\mathbf{z_i}=\mathbf{v_j}|i=H,\mathcal{H}_h)^{I(\mathbf{z_i},\mathbf{\mathbf{v_i}})}
\end{align}
for $h=0,1$, where 
\begin{equation}
    P(\mathbf{z_i}=\mathbf{v_j}|i=H,\mathcal{H}_h)=P(\tau_{i,j-1}\leq y_i\leq\tau_{i,j}|i=H,H_h)
\end{equation}
based on \eqref{eq:quantize} and $I(a,b)$ is an indicator function that returns 1 if $a$ equals $b$ and returns 0 otherwise.

Since $y_i|\mathcal{H}_0$ and $y_i|\mathcal{H}_1$ follow Gaussian distributions, we have
\begin{align}\label{eq:y_observ}
    &y_{i}|\mathcal{H}_0\sim\mathcal{N}(0,\beta_{i,0}^2)\\
    &y_{i}|\mathcal{H}_1\overset{a}{\sim}\mathcal{N}(0,\beta_{i,1}^2),\label{eq:y_i_h1}
\end{align}
respectively, where $\beta_{i,0}^2=\sigma_n^2$, $\beta_{i,1}^2=\sigma_n^2+p\sigma_x^2||\mathbf{h}_i||^2_2$ and $b\overset{a}{\sim}f(b)$ means variable $b$ asymptotically follows PDF $f(b)$. The proof of \eqref{eq:y_i_h1} is provided in [\cite{wimalajeewa2017compressive}, Appendix B], where the Lyapounov Central Limit Theorem (CLT) is utilized to derive the results. Let $A_{i,j,h}$ represent the probability that $y_i$ falls within the range of $[\tau_{i,j-1},\tau_{i,j}]$ when sensor $i$ is honest under hypothesis $\mathcal{H}_h$, i.e., $P(\tau_{i,j-1}\leq y_i\leq\tau_{i,j}|i=H,\mathcal{H}_h)$. Then $A_{i,j,h}$ is given by
\begin{equation}
    A_{i,j,h}=Q(\frac{\tau_{i,j-1}}{\beta_{i,h}})-Q(\frac{\tau_{i,j}}{\beta_{i,h}})   
\end{equation}
for $h=0,1$, where $Q(\cdot)$ denotes the tail distribution function of the standard normal distribution. 
\begin{figure}
    \centering
    \includegraphics[height=6em,width=8em]{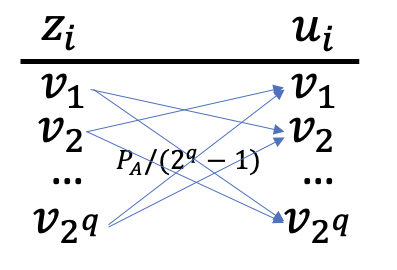}
    \caption{Attack model for a Byzantine node $i$. With a probability of $P_A/(2^q-1)$, each Byzantine node decides to send a decision that differs from the one it believes to be correct.}
    \label{fig:attack_model}
\end{figure}
If sensor $i$ is Byzantine, $\mathbf{u_i}$ does not have to be equal to $\mathbf{z_i}$. The attack model for Byzantine nodes is illustrated in Fig. \ref{fig:attack_model}. Thus, according to the chain rule, the PDF of local decision $\mathbf{u_i}$ is given as \eqref{eq:u_i_B}, where
\begin{align}
\label{eq:u=z}
P(\mathbf{u_i}=\mathbf{v_j}|\mathbf{u_i}=\mathbf{z_i},\mathbf{z_i}=\mathbf{v_k},i=B,\mathcal{H}_h)= 	\begin{cases}
		1&j=k\\
		0&j\neq k,
		\end{cases}
\end{align}
\begin{align}
\label{eq:u_z}
P(\mathbf{u_i}=\mathbf{v_j}|\mathbf{u_i}\neq \mathbf{z_i},\mathbf{z_i}=\mathbf{v_k},i=B,\mathcal{H}_h)= 	\begin{cases}
		0&j=k\\
		\frac{1}{2^q-1}&j\neq k,
		\end{cases}
\end{align}

\begin{figure*}[ht]
\begin{align}\label{eq:u_i_B}
    &P(\mathbf{u_i}|i=B,\mathcal{H}_h)=\prod_{j=1}^{2^q}P(\mathbf{u_i}=\mathbf{v_j}|i=B,\mathcal{H}_h)^{I(\mathbf{u_i},\mathbf{v_j})}\notag\\
    =&\prod_{j=1}^{2^q}[\sum_{k=1}^{2^q}P(\mathbf{z_i}=\mathbf{v_k}|i=B,\mathcal{H}_h)P(\mathbf{u_i}=\mathbf{z_i}|\mathbf{z_i}=\mathbf{v_k},i=B,\mathcal{H}_h) P(\mathbf{u_i}=\mathbf{v_j}|\mathbf{u_i}=\mathbf{z_i},\mathbf{z_i}=\mathbf{v_k},i=B,\mathcal{H}_h)\notag\\
    &+P(\mathbf{z_i}=\mathbf{v_k}|i=B,\mathcal{H}_h)P(\mathbf{u_i}\neq \mathbf{z_i}|\mathbf{z_i}=\mathbf{v_k},i=B,\mathcal{H}_h)P(\mathbf{u_i}=\mathbf{v_j}|\mathbf{u_i}\neq \mathbf{z_i},\mathbf{z_i}=\mathbf{v_k},i=B,\mathcal{H}_h)]^{I(\mathbf{u_i},\mathbf{v_j})}
\end{align}
\end{figure*}
\vspace{-2mm}
$P(\mathbf{u_i}\neq \mathbf{z_i}|\mathbf{z_i}=\mathbf{v_k},i=B,\mathcal{H}_h)=P_A$, $P(\mathbf{u_i}= \mathbf{z_i}|\mathbf{z_i}=\mathbf{v_k},i=B,\mathcal{H}_h)=1-P_A$ and $P(\mathbf{z_i}=\mathbf{v_k}|i=B,\mathcal{H}_h)=Q(\frac{\tau_{i,k-1}}{\beta_{i,h}})-Q(\frac{\tau_{i,k}}{\beta_{i,h}})$ for $h=0,1$. Note that  \eqref{eq:u=z} and \eqref{eq:u_z} are equivalent to $I(i,k)$ and $\frac{1-I(i,k)}{2^q-1}$, respectively.
Hence, \eqref{eq:u_i_B} can be rewritten as
\begin{small}
\begin{align}
    &P(\mathbf{u_i}|i=B,\mathcal{H}_h)\notag\\
    =&\prod_{j=1}^{2^q}\left\{\sum_{k=1}^{2^q}A_{i,k,h}\left[(1-P_A)I(j,k)+\frac{P_A(1-I(i,k))}{2^q-1}\right]\right\}^{I(\mathbf{u_i},\mathbf{v_j})}\notag\\
    =&\prod_{j=1}^{2^q}\left\{\sum_{k=1}^{2^q}A_{i,k,h}\left[(1-P_A-\frac{P_A}{2^q-1})I(j,k)+\frac{P_A}{2^q-1}\right]\right\}^{I(\mathbf{u_i},\mathbf{v_j})}\notag\\
    =&\prod_{j=1}^{2^q}\left\{A_{i,j,h}(1-P_A)+\sum_{k=1,k\neq j}^{2^q}A_{i,k,h}\frac{P_A}{2^q-1}\right\}^{I(\mathbf{u_i},\mathbf{v_j})}\notag\\
    =&\prod_{j=1}^{2^q}\left\{A_{i,j,h}(1-P_A)+(1-A_{i,j,h})\frac{P_A}{2^q-1}\right\}^{I(\mathbf{u_i},\mathbf{v_j})}\notag\\
    =&\prod_{j=1}^{2^q}P(\mathbf{u_i}=\mathbf{v_j}|i=B,\mathcal{H}_h)^{I(\mathbf{u_i},\mathbf{v_j})}.
\end{align}
\end{small}
Due to the statistical independence of the local decisions $\{u_1,u_2 ,\dots,u_N\}$, we have
\begin{align}\label{eq:u_attack}
    P(\mathbf{U}|\mathcal{H}_h)\!=\!\prod_{i=1}^N\!\prod_{j=1}^{2^q}\!\!\left[\!\!\sum_{X=B,H}\!\!\!\!P(\mathbf{u_i}\!=\!\mathbf{v_j}|i\!=\!X,\mathcal{H}_h)P(i\!=\!X)\!\right]^{I(\mathbf{u_i},\mathbf{v_j})}
\end{align}
for $h=0,1$.
\section{GLRT and Quantized LMPT Detectors}\label{sec:proposed}
In this section, we start with a brief review of the GLRT and the quantized LMPT detectors where all the sensors are assumed to be honest so that they send uncorrupted decisions to the FC, i.e., $\mathbf{u_i}=\mathbf{z_i}$. Then, the performance of the GLRT and the quantized LMPT detectors under Byzantine attacks is evaluated. The sparse signals here are characterized by the BG model. Under the BG model, the problem of distributed detection of sparse stochastic signals can be formulated as a problem of one-sided and close hypothesis testing which is given as
\begin{align}
\begin{cases}
    \mathcal{H}_0:&p=0\\
    \mathcal{H}_1:&p\rightarrow0^{+}.
    \end{cases}
\end{align}

\subsection{Fusion Rule for GLRT and Quantized LMPT Detectors with Honest Sensors}
\subsubsection{GLRT Detector}
The fusion rule of the GLRT detector is given by
\begin{align}\label{eq:FR_33}
    \frac{\max_{p}P(\mathbf{U}|\mathcal{H}_1;p)}{P(\mathbf{U}|\mathcal{H}_0;p=0)}\overset{\mathcal{H}_1}{\underset{\mathcal{H}_0}{\gtrless}}\lambda',
\end{align}
We can obtain the estimated sparsity degree $\hat{p}$ via maximum-likelihood estimation (MLE) which is given as $\hat{p}=\arg \max_{p} P(\mathbf{U}|\mathcal{H}_1;p)$. By replacing $p$ by $\hat{p}$ in \eqref{eq:FR_33} and taking the logarithm of both sides of \eqref{eq:FR_33}, the fusion rule can be expressed as
\begin{equation}\label{eq:quant_fr_glrt}
    \Gamma_{GLRT}=\sum_{i=1}^N\sum_{j=1}^{2^q}I(\mathbf{z_i}=\mathbf{v_j})g_{i,j}\overset{\mathcal{H}_1}{\underset{\mathcal{H}_0}{\gtrless}}\lambda_1,
\end{equation}
where $g_{i,j}=\hat{A}_{i,j,1}-\hat{A}_{i,j,0}$, $\hat{A}_{i,j,1}=Q(\frac{\tau_{i,j-1}}{\sqrt{\sigma_n^2+\hat{p}\sigma_x^2}})-Q(\frac{\tau_{i,j}}{\sqrt{\sigma_n^2+\hat{p}\sigma_x^2}})$ and $\hat{A}_{i,j,0}=A_{i,j,0}$.

\subsubsection{Quantized LMPT Detector}
Under $\mathcal{H}_1$, the sparsity degree $p$ is positive and close to zero,  and under $\mathcal{H}_0$, $p=0$. Hence, the problem of distributed detection of sparse stochastic signals can be performed via locally most powerful
tests as shown in \cite{wang2019distributed}.
Firstly, the logarithm form of the LRT, which is given by
\begin{align}\label{eq:lrT}
    lnP(\mathbf{U}|\mathcal{H}_1;p)-lnP(\mathbf{U}|\mathcal{H}_0)\overset{\mathcal{H}_1}{\underset{\mathcal{H}_0}{\gtrless}}ln(p_0/p_1),
\end{align}
is considered for decision-making at the FC,
where $P(\mathbf{U}|\mathcal{H}_h)=\prod_{i=1}^NP(\mathbf{u_i}|\mathcal{H}_h,i=H)$ and $P(\mathcal{H}_h)=p_h$ for $h=0,1$. Due to the fact that the sparsity degree $p$ is close to zero, the first-order Taylor's series expansion of $lnP(\mathbf{U}|\mathcal{H}_1;p)$ around zero is given as 
\begin{align}\label{eq:TAYLOR}
    lnP(\mathbf{U}|\mathcal{H}_1;p)\!=\!lnP(\mathbf{U}|\mathcal{H}_1;p\!=\!0)\!+\! p\left(\!\frac{\partial lnP(\mathbf{U}|\mathcal{H}_1;p)}{\partial p}\!\right)_{p\!=\!0}.
\end{align}
By substituting \eqref{eq:TAYLOR} in \eqref{eq:lrT}, the test statistic of the quantized LMPT detector is given by
\begin{equation}
    \left(\frac{\partial lnP(\mathbf{U}|\mathcal{H}_1;p)}{\partial p}\right)_{p=0}\overset{\mathcal{H}_1}{\underset{\mathcal{H}_0}{\gtrless}}\frac{ln(p_0/p_1)}{p}=\lambda_2,
\end{equation}
where
\begin{align}
    \frac{\partial lnP(\mathbf{U}|\mathcal{H}_1;p)}{\partial p}=&\sum_{i=1}^{N}\frac{\partial lnP(\mathbf{u_i}|\mathcal{H}_1,i=H;p)}{\partial p}\notag\\
    &=\sum_{i=1}^N\sum_{j=1}^{2^q}w_{i,j}I(\mathbf{u_i}=\mathbf{v_j})
\end{align}
and $w_{i,j}=\frac{\sigma^2_x||h_i||^2_2}{2\beta_{i,1}^{3}}\left[\tau_{i,j-1}\Phi(\frac{\tau_{i,j-1}}{\beta_{i,1}})-\tau_{i,j}\Phi(\frac{\tau_{i,j}}{\beta_{i,1}})\right]A_{i,j,1}^{-1}$. Here, $\Phi(\cdot)$ denotes the cumulative distribution function (CDF) of the standard normal distribution. Hence, the decision rule is given as
\vspace{-2mm}
\begin{equation}\label{eq:quant_fr}
    \Gamma_{LMPT}=\sum_{i=1}^N\sum_{j=1}^{2^q}I(\mathbf{u_i}=\mathbf{v_j})\widetilde{w}_{i,j}\overset{\mathcal{H}_1}{\underset{\mathcal{H}_0}{\gtrless}}\lambda_2,
\end{equation}
\vspace{-3mm}
where $\widetilde{w}_{i,j}=\left(w_{i,j}\right)_{p=0}$.
\subsection{Performance Analysis of the GLRT and the Quantized LMPT Detectors in the Presence of Byzantines}
In this subsection, we evaluate the detection performance of the GLRT and the quantized LMPT detectors in the presence of Byzantines. We also derive the optimal attack strategy of the Byzantines.

Let $L=\sum_{i=1}^NL_i$ denote the global statistic for the fusion rule given in \eqref{eq:quant_fr_glrt} or \eqref{eq:quant_fr}, where $L_i=\sum_{j=1}^{2^q}I(\mathbf{u_i}=\mathbf{v_j})d_{i,j}$ and $d_{i,j}\in\left\{\widetilde{w}_{i,j},g_{i,j}\right\}$. According to the Lyapunov CLT, $L$ approximately follows a Gaussian distribution with mean $E(\sum_{i=1}^NL_i)$ and variance $Var(\sum_{i=1}^NL_i)$ when $N$ is sufficiently large. Under both hypotheses, $E(L)$ and $Var(L)$ are given as
\begin{align}\label{eq:mean_L}
    E(L|\mathcal{H}_h)=&\sum_{i=1}^NE(L_i|\mathcal{H}_h)=\sum_{i=1}^NE\left(\sum_{j=1}^{2^q}I(\mathbf{u_i}=\mathbf{v_j})d_{i,j}\right)\notag\\
    =&\sum_{i=1}^N\sum_{j=1}^{2^q}P(\mathbf{u_i}=\mathbf{v_j}|\mathcal{H}_h)d_{i,j}\notag\\
    =&\sum_{i=1}^N\sum_{j=1}^{2^q}[P(\mathbf{u_i}=\mathbf{v_j}|\mathcal{H}_h,i=H)(1-\alpha)\notag\\
    &+P(\mathbf{u_i}=\mathbf{v_j}|\mathcal{H}_h,i=B)\alpha]d_{i,j}
\end{align}
and
\begin{align}\label{eq:var_L}
    Var(L|\mathcal{H}_h)=&\sum_{i=1}^NVar(L_i|\mathcal{H}_h)=\sum_{i=1}^N\left[E\left(L_i^2|\mathcal{H}_h\right)-E(L_i|\mathcal{H}_h)^2\right]\notag\\
    =&\sum_{i=1}^NE\left[\left(\sum_{j=1}^{2^q}I(\mathbf{u_i}=\mathbf{v_j})d_{i,j}\right)^2\right]-E(L|\mathcal{H}_h)^2\notag\\
    =&\sum_{i=1}^N\sum_{j=1}^{2^q}P(\mathbf{u_i}=\mathbf{v_j}|\mathcal{H}_h)d_{i,j}^2-E(L|\mathcal{H}_h)^2\notag\\
    =&\sum_{i=1}^N\sum_{j=1}^{2^q}[P(\mathbf{u_i}=\mathbf{v_j}|\mathcal{H}_h,i=H)(1-\alpha)\notag\\
    &+P(\mathbf{u_i}=\mathbf{v_j}|\mathcal{H}_h,i=B)\alpha]d_{i,j}^2-E(L|\mathcal{H}_h)^2,
\end{align}
respectively. Using the expression in \eqref{eq:mean_L} and \eqref{eq:var_L}, the probabilities of detection and false alarm can be calculated as
\begin{align}
    P_d=P(L>\lambda|\mathcal{H}_1)=Q\left(\frac{\lambda-E(L|\mathcal{H}_1)}{\sqrt{ Var(L|\mathcal{H}_1)}}\right)
\end{align}
and
\begin{align}
    P_f=P(L>\lambda|\mathcal{H}_0)=Q\left(\frac{\lambda-E(L|\mathcal{H}_0)}{\sqrt{ Var(L|\mathcal{H}_0)}}\right),
\end{align}
respectively, where $\lambda\in\{\lambda_1,\lambda_2\}$.

From the perspective of attackers, the optimal attack strategy is to make the system blind, i.e., to make the probability of detection equal to 1/2. The deflection coefficient is utilized here as a useful tool to determine the optimal attack strategy due to its simplicity and strong relationship with the global probability of detection. The deflection coefficient is defined as $D_f=\frac{(E(L|\mathcal{H}_1)-E(L|\mathcal{H}_0))^2}{Var(L|\mathcal{H}_1)}$.
To blind the FC, Byzantines strategically design the attack parameters so that $D_f=0$, i.e., $E(L|\mathcal{H}_1)=E(L|\mathcal{H}_0)$. By utilizing \eqref{eq:mean_L}, we can obtain
\begin{align}\label{eq:blind}
    \alpha P_A=\frac{\sum_{i=1}^N\sum_{j=1}^{2^q}(A_{i,j,1}-A_{i,j,0})d_{i,j}}{\sum_{i=1}^N\sum_{j=1}^{2^q}\left[\frac{1}{2^q-1}+(1-\frac{1}{2^q-1})(A_{i,j,1}-A_{i,j,0})\right]d_{i,j}}.
\end{align}
Thus, we can conclude that the attackers are able to blind the FC when $\alpha P_A$ equals the right-hand side of \eqref{eq:blind}. From the simulation results presented later in Sec. \ref{sec:simul}, both the GLRT and the quantized LMPT detectors are very vulnerable to Byzantine attacks, even if the attack parameter $P_A$ is very small. A possible explanation could be that, since detectors make their decisions based on observations with the same mean and slightly different variances under the two hypotheses, it is easy for them to make incorrect decisions in the presence of Byzantines.

\section{Robust fusion rule}\label{sec:proposed2}
In order to improve the robustness of the detector, we attempt to elicit some additional information regarding the attack parameters from the local decisions of some sensors and incorporate it into the design of the fusion rule. In general, a detector's performance improves as additional information is obtained, e.g., sparsity degree $p$, the fraction of Byzantines $\alpha$, and attack probability $P_A$. Intuitively, a GLRT detector can be designed, which takes both the unknown sparsity degree and the unknown attack parameters into consideration, as shown in \eqref{eq:FR_333}.
\begin{align}\label{eq:FR_333}
    \frac{\max_{p,P_A,\alpha}P(\mathbf{U}|\mathcal{H}_1;p)}{\max_{P_A,\alpha}P(\mathbf{U}|\mathcal{H}_0;p=0)}\overset{\mathcal{H}_1}{\underset{\mathcal{H}_0}{\gtrless}}\lambda''.
\end{align}
Usually, suppose the sparse signals are weak. In that case, the number of sensors is large,  the MLE attains its asymptotic PDF, and an appropriate threshold $\lambda''$  can be found based on the asymptotic detection performance of the GLRT detectors (see Sec. 6.5 in \cite{kay1993fundamentals}). However, sparse signals do not necessarily indicate weak signals. Thus, it is not always tractable to obtain an appropriate threshold value $\lambda''$. Moreover, the presence of nuisance parameters $P_A$ and $\alpha$ results in a degradation of the detection performance of GLRT detectors.

To overcome these problems, as alluded to earlier, we randomly select a fraction of the sensors as reference sensors from the set of all sensors and estimate unknown parameters (i.e., $\alpha$, $P_A$ and $p$) in two steps.
In the first step, nuisance attack parameters are estimated based on the local decisions coming from reference sensors. In the second step, the estimated attack parameters are utilized to estimate the unknown sparsity degree $p$ based on the local decisions from the remaining sensors. The proposed GLRTRS detector is based on the above parameter estimates. As the LMPT-based detector does not require the knowledge of the sparsity degree $p$, the only estimation occurs in the first step, which is the estimation of the nuisance attack parameters. Later in this section, we will provide details about the proposed GLRTRS and LMPTRS detectors.

Since we carry out the entire estimation process in two steps, we would like to minimize the performance loss caused by partitioning the estimation process. Let us take the GLRT detector presented in \eqref{eq:FR_333} as an example. Suppose we want to partition the entire estimation process into two steps, as described above. In that case, we want to ensure that the performance degradation caused by the unknown sparsity degree $p$ is negligible while estimating the attack parameters. In other words, the two pairs of estimated attack parameters we obtain, which are $\{\alpha_{H_1},P_{A,H_1}\}=\arg \max_{\alpha,P_A} P(\mathbf{U}|\mathcal{H}_1,p,\alpha,P_A)$ and $\{\alpha_{H_0},P_{A,H_0}\}=\arg \max_{\alpha,P_A} P(\mathbf{U}|\mathcal{H}_0,p=0,\alpha,P_A)$, should be very close to each other. To complete this task, we introduce reference sensors to assist us. We randomly and uniformly select a set of reference sensors from the set of all the sensors to specifically estimate the unknown nuisance attack parameters $P_A$ and $\alpha$.\footnote{Since we have assumed that $\alpha$ fraction of Byzantine nodes are uniformly distributed in the network, there are $\alpha$ fraction of Byzantine nodes within both the set of reference sensors and remaining sensors.} At the reference sensors, we employ different predefined thresholds so that the decisions of the reference sensors satisfy Assumption 1 below.

{\em Assumption 1:} $Pr(\mathbf{z_i}=v_{2^q}|\mathcal{H}_1)\rightarrow{1}$ and $Pr(\mathbf{z_i}=v_{2^q}|\mathcal{H}_0)\rightarrow{1}$ (or $Pr(\mathbf{z_i}=v_{1}|\mathcal{H}_1)\rightarrow{1}$ and $Pr(\mathbf{z_i}=v_{1}|\mathcal{H}_0)\rightarrow{1}$).

It is implied by Assumption 1 that the quantizer thresholds $\{\widetilde\tau_{i,j}\}_{j=1}^{2^q}$ for any reference sensor $i$ are designed in such a manner that sensor $i$ sends the same decisions under both hypotheses with probabilities that approach 1. One possible and simplest method is to give the reference sensors lower (or higher) predefined quantizer thresholds compared with other sensors, i.e., $\widetilde\tau_{j,2^q-1}\ll\tau_{i,1}$ (or $\tau_{i,2^q}\ll\widetilde\tau_{j,1}$).\footnote{Based upon \eqref{eq:y_observ}, the observation $y_i$ for $i\in\{1,2,\dots,N\}$ has zero mean and different variances that are related to sparsity degree $p$ given different hypotheses. Since a sparse signal is considered in the paper for which the sparsity degree $p$ tends to 0, it is possible to design reasonable quantizer thresholds for reference nodes. A reasonable quantizer threshold refers to a quantizer threshold that is not excessively large or small. From experiments, it has been shown that $\tau_{i,1}-\widetilde\tau_{j,2^q-1}=6$ (or $\widetilde\tau_{j,1}-\tau_{i,2^q}=6$) is sufficient to satisfy Assumption 1 for reference sensors.} If Assumption 1 is satisfied, the reference sensors continue to send the same decision regardless of the true underlying hypothesis. It allows us to ensure that the performance degradation caused by the unknown sparsity degree $p$ is negligible while the attack parameters are being estimated.

In the following subsections, we consider two cases: (i) The sparsity degree $p$ and the attack parameters $\{\alpha,P_A\}$ are all unknown; (ii) $\alpha$ is known, both sparsity degree $p$ and attack probability $P_A$ are unknown. 

\subsection{Networks with Unknown $p$, $\alpha$ and $P_A$}
We propose two detectors in this subsection: the GLRTRS detector that requires the estimation of unknown parameter $p$, and the LMPTRS detector that does not require the estimation of $p$.

\paragraph{GLRTRS detector}
According to \eqref{eq:u_attack}, we are able to obtain
\begin{align}
    \!P(\mathbf{U}|\mathcal{H}_h)\!=\!\prod_{i=1}^N\!\prod_{j=1}^{2^q}\!\left[\!A_{i,j,h}\!+\!x\left(\frac{1}{2^q\!-\!1}\!-\!A_{i,j,h}\!-\!\frac{A_{i,j,h}}{2^q\!-\!1}\right)\!\right]^{I(\mathbf{u_i},\mathbf{v_j})}
\end{align}
where $x=\alpha P_A$. For convenience, instead of considering the two attack parameters $\alpha$ and $P_A$ separately, we consider a single attack parameter $x$. The problem of distributed detection of a sparse stochastic signal can be formulated as
\begin{align}
\begin{cases}
    \mathcal{H}_0:&p=0, 0\leq x \leq 1\\
    \mathcal{H}_1:&p\rightarrow0^{+}, 0\leq x\leq 1
    \end{cases}.
\end{align}
The fusion rule of the GLRTRS detector is given by
\begin{align}\label{eq:FR_11}
    \frac{\max_{p,x}\prod_{i=N_{ref}+1}^NP(\mathbf{u_i}|\mathcal{H}_1,p,x)}{\max_{x}\prod_{i=N_{ref}+1}^NP(\mathbf{u_i}|\mathcal{H}_0,p=0,x)}\overset{\mathcal{H}_1}{\underset{\mathcal{H}_0}{\gtrless}}\lambda,
\end{align}
where $N_{ref}$ is the number of reference sensors and they are labelled as $1, 2,3\ldots, N_{ref}$.
We first utilize the data from the reference sensors to estimate the unknown attack parameter $x$ via MLE. The estimated attack parameter $x$ is given as
\begin{align}\label{eq:MLE}
    x_{H_h}=\arg \max_{x} P(\mathbf{U}_{ref}|\mathcal{H}_h,p,x)
\end{align}
for $h=0,1$. $P(\mathbf{U}_{ref}|\mathcal{H}_h,p,x)$ in \eqref{eq:MLE} is the joint pmf of local decisions coming from the reference sensors and it is given as
\begin{align}
    &P(\mathbf{U}_{ref}|\mathcal{H}_h,p,x)\notag\\
    \!=&\!\prod_{i=1}^{N_{ref}}\!\prod_{j=1}^{2^q}\!\!\left[\!\!\sum_{X=B,H}\!\!\!\!P(\mathbf{u_i}\!=\!\mathbf{v_j}|i\!=\!X,\mathcal{H}_h)P(i\!=\!X)\!\right]^{I(\mathbf{u_i},\mathbf{v_j})}\notag\\
    \!=&\!\prod_{i=1}^{N_{ref}}\!\prod_{j=1}^{2^{q}}\!\!\left[C_{i,j,h}\!+\!x\left(\frac{1}{2^q\!-\!1}\!-\!C_{i,j,h}\!-\!\frac{1}{2^q\!-\!1}C_{i,j,h}\right)\right]^{I(\mathbf{u_i}=\mathbf{v_j})}
\end{align}
for $h=0,1$, where $C_{i,j,h}=Q(\frac{\widetilde\tau_{i,j-1}}{\beta_{i,h}})-Q(\frac{\widetilde\tau_{i,j}}{\beta_{i,h}})$. By replacing $x$ by $x_{H_1}$ in $P(\mathbf{u_i}|\mathcal{H}_1,p,x_{H_1})$ and $x$ by $x_{H_0}$ in $P(\mathbf{u_i}|\mathcal{H}_1,p=0,x_{H_0})$ in \eqref{eq:FR_11}, the new fusion rule is given by
\begin{align}\label{eq:FR_1}
    \frac{\max_{p}\prod_{i=N_{ref}+1}^NP(\mathbf{u_i}|\mathcal{H}_1,p,x_{H_1})}{\prod_{i=N_{ref}+1}^NP(\mathbf{u_i}|\mathcal{H}_0,p=0,x_{H_0})}\overset{\mathcal{H}_1}{\underset{\mathcal{H}_0}{\gtrless}}\kappa,
\end{align}
where $P(\mathbf{u_i}|\mathcal{H}_h,p,x_{H_h})=\prod_{j=1}^{2^{q}}P(\mathbf{u_i}=\mathbf{v_j}|\mathcal{H}_h,p,x_{H_h})$. We have the following theorem stating that the estimator in \eqref{eq:MLE} is an efficient MLE when Assumption 1 is satisfied.
\begin{theorem}
The ML estimation of the unknown attack parameter $x$ based on the data from the reference sensors is unbiased, and it attains the CRLB of the problem, which equals $\frac{(1-x)x}{N_{ref}}$.
\end{theorem}
\begin{IEEEproof}
 Please see Appendix \ref{Proof_theo}. 
\end{IEEEproof}
Since we estimate the attack parameter $x$ based on observations from the reference sensors, $x_{H_0}$ is approximately the same as $x_{H_1}$, i.e., $x_{H_0}\approx x_{H_1}$ (according to Assumption 1). Let $x_{H}=\frac{x_{H_1}+x_{H_0}}{2}$ represent the averaged estimated $x$ and replace both $x_{H_1}$ and $x_{H_0}$ with $x_{H}$ in $\eqref{eq:FR_1}$, the fusion rule can be expressed as
\begin{align}\label{eq:FR_2}
    \frac{\prod_{i=N_{ref}+1}^NP(\mathbf{u_i}|\mathcal{H}_1,p,x_H)}{\prod_{i=N_{ref}+1}^NP(\mathbf{u_i}|\mathcal{H}_0,p=0,x_H)}\overset{\mathcal{H}_1}{\underset{\mathcal{H}_0}{\gtrless}}\kappa,
\end{align}
where $\kappa$ is the threshold to be set in order to ensure the desired probability of false alarm PFA. Next, we calculate the estimated sparsity degree $\hat{p}$, which is given as $\hat{p}=\arg\max_{p}\prod_{i=N_{ref}+1}^NP(\mathbf{u_i}|\mathcal{H}_1,p,x_{H})$. Upon taking the logarithm of both sides of \eqref{eq:FR_2}, the simplified fusion rule is given as
\begin{equation}\label{FR:glrtrs}
    \Gamma_{GLRTRS}=\sum_{i=N_{ref}+1}^{N}\sum_{j=1}^{2^{q}}{I(\mathbf{u_i}=\mathbf{v_j})}F_{i,j}\overset{\mathcal{H}_1}{\underset{\mathcal{H}_0}{\gtrless}}\kappa',
\end{equation}
where $\kappa'=\log(\kappa)$, $F_{i,j}=f_{i,j,1}-f_{i,j,0}$, $f_{i,j,h}=\hat{A}_{i,j,h}+x_H\left(\frac{1}{2^q-1}-\hat{A}_{i,j,h}-\frac{1}{2^q-1}\hat{A}_{i,j,h}\right)$, $\hat{A}_{i,j,1}=Q(\frac{\tau_{i,j-1}}{\sqrt{\sigma_n^2+\hat{p}\sigma_x^2}})-Q(\frac{\tau_{i,j}}{\sqrt{\sigma_n^2+\hat{p}\sigma_x^2}})$ and $\hat{A}_{i,j,0}=A_{i,j,0}$. Assume that $N-N_{Nef}$ is sufficiently large, the global statistic $\Gamma_{GLRTRS}$ then follows a Gaussian distribution with mean
\begin{align}\label{eq:meanmean}
    E(\Gamma_{GLRTRS}|H_h)\!\!=\!\!\!\!\!\sum_{i=N_{ref}+1}^{N}\!\sum_{j=1}^{2^{q}}F_{i,j}P(\mathbf{u_i}=\mathbf{v_j}|H_h,x_H,p)
\end{align}
and variance
\begin{small}
\begin{align}\label{eq:varvar}
    Var(\Gamma_{GLRTRS}|H_h)=&\!\!\!\sum_{i=N_{ref}+1}^{N}\!\sum_{j=1}^{2^{q}}F_{i,j}^2P(\mathbf{u_i}=\mathbf{v_j}|H_h,x_H,p)\notag\\
    &-E^2(\Gamma_{GLRTRS}|H_h)
\end{align}
\end{small}
for $h=0,1$. With \eqref{eq:meanmean} and \eqref{eq:varvar}, the probabilities of detection and false alarm are respectively given as
\begin{align}
    P_d=&Q\left(\frac{\kappa'-E(\Gamma_{GLRTRS}|H_1)}{\sqrt{Var(\Gamma_{GLRTRS}|H_1)}}\right),\\
    P_f=&Q\left(\frac{\kappa'-E(\Gamma_{GLRTRS}|H_0)}{\sqrt{Var(\Gamma_{GLRTRS}|H_0)}}\right).
\end{align}
For a given false alarm $PFA$, we can obtain the suboptimal adaptive threshold used by the FC as shown in \eqref{eq:thre1}.\footnote{Since we obtain the adaptive threshold based on the estimated attack parameter, it is a suboptimal threshold that approximately satisfies a desired false alarm.}
\begin{align}\label{eq:thre1}
    \kappa'\!=\!Q^{-1}\!\left(PFA\right)\!\sqrt{Var(\Gamma_{GLRTRS}|H_0)}\!+\!E(\Gamma_{GLRTRS}|H_0)
\end{align}

\paragraph{LMPTRS detector}
Similarly, after we obtain the estimated attack parameter $x_H$, the test statistic of the proposed LMPTRS detector can be expressed as
\begin{equation}
    \left(\frac{\partial lnP(\mathbf{U}|\mathcal{H}_1,p,x_H)}{\partial p}\right)_{p=0}\overset{\mathcal{H}_1}{\underset{\mathcal{H}_0}{\gtrless}}\frac{ln(p_0/p_1)}{p},
\end{equation}
where
\begin{align}
    &\frac{\partial lnP(\mathbf{U}|\mathcal{H}_1,p,x_H)}{\partial p}=\sum_{i=1}^{N}\frac{\partial lnP(\mathbf{u_i}|\mathcal{H}_1,p,x_H)}{\partial p}\notag\\
    =&\!\sum_{i=1}^N\!\sum_{j=1}^{2^q}\frac{\sigma^2_x||h_i||^2_2I(\mathbf{u_i}\!=\!\mathbf{v_j})}{2(p\sigma^2_x||h_i||^2_2\!+\!\sigma^2_n)^{\frac{3}{2}}}\!\left[\!\tau_{i,j-1}\Phi(\frac{\tau_{i,j-1}}{\sqrt{p\sigma^2_x||h_i||^2_2+\sigma^2_n}})\right.\notag\\
    &\left.\!-\!\tau_{i,j}\Phi(\!\frac{\tau_{i,j}}{\sqrt{p\sigma^2_x||h_i||^2_2\!+\!\sigma^2_n}})\!\right]\!\frac{1-x_H-x_HA_{i,j,1}}{A_{i,j,1}\!+\!x_H(1\!-\!x_H\!-\!x_HA_{i,j,1})}\notag\\
    =&\sum_{i=1}^N\sum_{j=1}^{2^q}I(\mathbf{u_i}=\mathbf{v_j})g_{i,j}.
\end{align}
The fusion rule can be reformulated as
\begin{equation}\label{eq:similar_procedure3}
    \Gamma_{LMPTRS}=\sum_{i=1}^N\sum_{j=1}^{2^q}I(\mathbf{u_i}=\mathbf{v_j})\widetilde{g}_{i,j}\overset{\mathcal{H}_1}{\underset{\mathcal{H}_0}{\gtrless}}\gamma',
\end{equation}
where $\gamma'=\frac{ln(p_0/p_1)}{p}$ and $\widetilde{g}_{i,j}=(g_{i,j})_{p=0}$. Like the one employed earlier, we can derive the threshold $\gamma'$ in \eqref{eq:similar_procedure3} for a given false alarm $PFA$. We can obtain that $\gamma'=Q^{-1}\left(PFA\right)\sqrt{Var(\Gamma_{LMPTRS}|H_0)}+E(\Gamma_{LMPTRS}|H_0)$ , where $E(\Gamma_{LMPTRS}|H_0)=\sum_{i=N_{ref}+1}^{N}\sum_{j=1}^{2^{q}}\widetilde{w}_{i,j}P(\mathbf{u_i}=\mathbf{v_j}|H_0,x_H,p=0)$ and $Var(\Gamma_{LMPTRS}|H_0)=\sum_{i=N_{ref}+1}^{N}\sum_{j=1}^{2^{q}}\widetilde{w}_{i,j}^2P(\mathbf{u_i}=\mathbf{v_j}|H_0,x_H,p=0)-E^2(\Gamma_{LMPTRS}|H_0)$.

\subsection{Networks with Known $\alpha$, Unknown $p$ and Unknown $P_A$}
When it is assumed that we know the fraction of Byzantine nodes $\alpha$ in the network, we can obtain more accurate information and achieve better detection performance. In this subsection, the GLRTRS and the LMPTRS detectors are further enhanced by introducing a local decision filter at the FC, which allows us to select sensors that are more likely to be honest. The proposed enhanced detectors are referred to as the E-GLRTRS and the E-LMPTRS detectors.

Upon receiving local decisions $\{\mathbf{U}(1),\dots,\mathbf{U}(t)\}$ until time step $t$, where $\mathbf{U}(t)=\{u_{1}(t),\dots,u_{N}(t)\}$, each sensor's statistical behavior is used to filter local decisions. The local decision filter distinguishes malicious nodes from honest nodes at time $t$ by the following
\begin{equation}\label{eq:filter}
    \sum_{j=1}^{2^q}\!|R_j\!-\!\widetilde{p}_t(\mathbf{u_i}\!=\!\mathbf{v_j})|\overset{b_i(t)=1}{\underset{b_i(t)=0}{\gtrless}}\tau, \forall{i}\in\{N_{ref}+1,\dots,N\},
\end{equation}
where $R_j\!=\!min\!\left(\!P(\!\mathbf{u_i}\!=\!\mathbf{v_j}|i\!=\!H,\mathcal{H}_1\!),P(\!\mathbf{u_i}\!=\!\mathbf{v_j}|i\!=\!H,\mathcal{H}_0\!)\!\right)$ is a benchmark value to filter out the potential malicious sensors \footnote{Note that based upon \eqref{eq:y_observ}, the observation $y_i, \forall{i}\in\{1,2,\dots,N\}$ has zero mean and different variances that are related to the sparsity degree $p$ given different hypotheses. Regardless of the quantizer thresholds that have been chosen, sensors tend to transmit the same decisions with slightly different probabilities based upon different hypotheses, i.e, $P(\mathbf{u_i}=\mathbf{v_j}|i=H,\mathcal{H}_1)$ and $P(\mathbf{u_i}=\mathbf{v_j}|i=H,\mathcal{H}_0)$ are slightly different. The simplest method of choosing $R_j$ is to take the minimum value between $P(\mathbf{u_i}=\mathbf{v_j}|i=H,\mathcal{H}_1)$ and $P(\mathbf{u_i}=\mathbf{v_j}|i=H,\mathcal{H}_0)$.} and $b_i(t)$ represents the behavioral identity of sensor $i$ at time $t$. If $b_i(t)=1$, the sensor $i$ is regarded as an honest node; otherwise, it is regarded as a potential Byzantine node. $\widetilde{p}_t(\mathbf{u_i}=\mathbf{v_j})$ is the empirical probability of $\mathbf{u_i}=\mathbf{v_j}$ until time step $t$ according to the history of local decisions and it is given as
\begin{equation}
    \widetilde{p}_t(\mathbf{u_i}=\mathbf{v_j})=\frac{\sum_{q=1}^tI(\mathbf{u_i}(t),\mathbf{v_j})}{t},
\end{equation}
where $\mathbf{u_i}(t)$ is the $\mathbf{u_i}$ at time step $t$.
The left side of \eqref{eq:filter} measures the deviation of the empirical probability of $\mathbf{u_i}=\mathbf{v_j}$ from the benchmark value $R_j$. Sensors are potential Byzantine nodes if the deviation exceeds a predefined threshold  $\tau$.
Based on the 
behavioral identity of all the sensors $\{b_i(t)\}_{i=1}^N$ at time step $t$, we can obtain the fusion rules of enhanced detectors. Note that GLRTRS and LMPTRS both have the form
\begin{align}
    \sum_{i=N_{ref}+1}^N\sum_{j=1}^{2^q}I(\mathbf{u_i}=\mathbf{v_j})W_{i,j}\overset{\mathcal{H}_1}{\underset{\mathcal{H}_0}{\gtrless}}\eta,
\end{align}
where $(W_{i,j},\eta)\in\{(\widetilde{g}_{i,j},\gamma'),(F_{i,j},\kappa')\}$. Hence, the enhanced fusion rule at time step $t$ is given by
\begin{align}\label{FR:enhanced}
    \Gamma_E(t)=\!\!\!\!\sum_{i=N_{ref}+1}^N\!\sum_{j=1}^{2^q}b_i(t)I(\mathbf{u_i}(t)\!=\!\mathbf{v_j})W_{i,j}(t)\overset{\mathcal{H}_1}{\underset{\mathcal{H}_0}{\gtrless}}\eta(t).
\end{align}
Let $\alpha_t(t)$ and $P_{A}(t)$ denote the probability that a sensor is a Byzantine node and the probability that a Byzantine node attacks at time step $t$, respectively, and $\alpha_t(0)=\alpha$ is the initial value of $\alpha_t$. We first obtain the estimated $\hat{p}_{A}(0)=x_H(0)/\alpha_t(0)$ given a prior knowledge of $\alpha_t$, i.e., $\alpha_0$, as initial values of $\hat{p}_{A}$, where $x_{H}(0)=\frac{x_{H_1}(0)+x_{H_0}(0)}{2}$ and $x_{H_h}(0)$ is given in \eqref{eq:MLE} for $h=0,1$. After filtering the possible Byzantine nodes, the value of $\alpha_t$ at time step $t=0$ is updated according to $\{b_i(0)\}_{i=N_{ref}+1}^N$. The updating rule is given as
\begin{align}\label{eq:update_alpha}
    \alpha_t(0)=\alpha-\frac{\sum_{i=N_{ref}+1}^Nb_i(0)}{N-N_{ref}}.
\end{align}

At the next time step, the updated $\alpha_t(0)$ is employed as a new prior to estimate $\hat{p}_{A}(1)$. The value of $\alpha_t$ is also updated at time step $t=1$ according to $\{b_i(1)\}_{i=N_{ref}+1}^N$ in the same manner as \eqref{eq:update_alpha}, i.e., $\alpha(1)=\alpha(0)-\frac{\sum_{i=N_{ref}+1}^Nb_i(1)}{N-N_{ref}}$, and becomes the new prior at the next time step. Thus, at time step $t$, $\alpha_t(t-1)=\alpha_t(t-2)-\frac{\sum_{i=N_{ref}+1}^Nb_i(t-1)}{N-N_{ref}}$ is utilized to estimate $\hat{p}_{A}(t)$ and $x_H(t)=\hat{p}_{A}(t)\alpha_t(t-1)$. By replacing $x_H$ and $F_{i,j}$ with $x_H(t)$ and $b_i(t)W_{i,j}$, respectively, in \eqref{eq:meanmean} and \eqref{eq:varvar}, we can obtain $E(\Gamma_{E}(t)|H_h)$ and $Var(\Gamma_{E}(t)|H_h)$. Similarly, for a given false alarm $PFA$, the threshold used by the FC at time step $t$ is given as $\eta(t)=Q^{-1}\left(PFA\right)\sqrt{Var(\Gamma_{E}(t)|H_0)}+E(\Gamma_{E}(t)|H_0)$.

\section{Simulation results}\label{sec:simulation}
\label{sec:simul}
 In this section, we present the simulation results to evaluate the performance of the proposed detectors in the presence of Byzantine attacks and compare them with the quantized LMPT detector (proposed in \cite{wang2019distributed}) and the commonly used GLRT detector. Via simulations, we analyze the performance of the proposed schemes in terms of the probability of error in the system. The linear operators $\{\mathbf{h_i}\}_{i=1}^{N}$ are all assumed to be sampled from normal distribution with a homogeneous scenario so that $||\mathbf{h_i}||_2=1,\forall i$ as described in \cite{wang2019distributed}. We set $\sigma^2_n=1$, $\sigma^2_x=5$, $PFA=0.4$ and $\alpha=0.3$. For all experiments, $\pi_1=\pi_0=0.5$. Unless otherwise noted, we assume the number of sensors $N$ to be 280. When reference sensors are employed, we employ $N_{ref}=80$ out of 280 sensors as reference sensors, except when we evaluate system performance as a function of $N_{ref}$. 

\begin{figure}
    \centering
    \includegraphics[height=15em,width=22em]{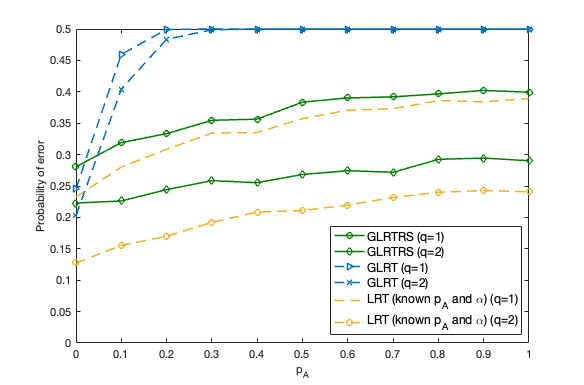}
    \caption{
    Comparison of $Pe$ for the GLRTRS, LRT and GLRT detectors.}
    \label{fig:fig3}
\end{figure}

In Fig.~\ref{fig:fig3}, we demonstrate the error probabilities of the LRT detector with perfect knowledge of $\{P_A, \alpha, p\}$, the GLRT detector, and the proposed GLRTRS detector. Two different quantizers are employed, i.e., $q=1$ and $q=2$. The error probability of the LRT detector with perfect knowledge of $\{P_A, \alpha, p\}$ shown in Fig.~\ref{fig:fig3} is used as the benchmark to assess the performance of the proposed detectors. It can be observed that the GLRT detector is extremely vulnerable to attacks for both one-bit quantization and multilevel quantization, and a small fraction of Byzantine nodes $\alpha$ with a small attack parameter $P_A$ are sufficient to break down the entire system. However, the proposed GLRTRS detector can obtain an error probability close to that of the LRT detector with perfect knowledge of $\{P_A, \alpha, p\}$. We can observe from Fig.~\ref{fig:fig3} that in the cases of $q=1$ and $q=2$, the GLRTRS detector outperforms the commonly used GLRT detector, with a performance close to the benchmark LRT detector. Note that the GLRTRS detector uses only 200 sensors for detection purposes and exhibits performance close to the benchmark detector that uses 280 sensors for detection purposes. 
With an increase in $q$, the error probability of the proposed GLRTRS detector further decreases due to the reduction of performance losses caused by quantization. From Fig.~\ref{fig:fig3}, we can also observe that the difference between the benchmark error probability and the error probability of the proposed GLRTRS detector is larger when the value of $q$ increases. It is because the GLRTRS detector is a sub-optimal detector, while the benchmark LRT detector is an optimal one. 
\begin{figure}
    \centering
    \includegraphics[height=15em,width=22em]{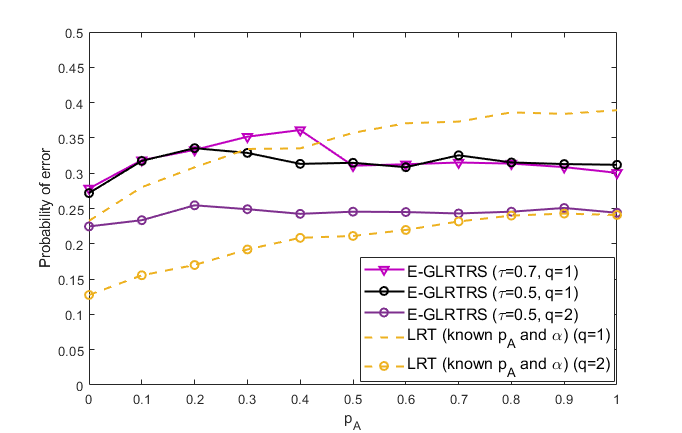}
    \caption{
    $Pe$ vesus $P_A$ when different values of $q$ and the different values of threshold $\tau$ are utilized for the E-GLRTRS detectors.}
    \label{fig:fig4}
\end{figure}

If we assume that the fraction of Byzantine nodes $\alpha$ is known to the system, The error probability of the system can be further reduced by employing the E-GLRTRS detector. As shown in Fig.~\ref{fig:fig4}, the error probability of the E-GLRTRS detector decreases with an appropriately designed threshold $\tau$ compared to the GLRTRS detector. We can filter out different numbers of potential Byzantine nodes with different values of the threshold $\tau$ in \eqref{eq:filter}. A potential Byzantine node can be either an actual Byzantine or a falsely identified one. It is obvious that a smaller threshold results in greater false filtering, while a larger threshold results in greater miss filtering. False filtering implies that honest nodes are falsely filtered out, whereas miss filtering implies that malicious nodes remain unfiltered. Both false filtering and miss filtering result in degrading the system's performance. Therefore, the system will likely perform better if the threshold $\tau$ is set appropriately. As shown in Fig.~\ref{fig:fig4}, $\tau=0.5$ is more appropriate than $\tau=0.7$. It can be observed that when $\tau=0.5$, $q=1$ and $P_A>0.3$, the E-GLRTRS detector outperforms the LRT detector with perfect knowledge of $\{P_A, \alpha, p\}$. This is because the E-GLRTRS detector filters out potential Byzantine nodes and utilizes the rest of the sensors for detection. In contrast, the benchmark LRT detector utilizes all the sensors for detection purposes. However, the E-GLRTRS detector is inferior to the benchmark LRT detector when $q=1$ and $P_A<0.3$, the difference in error probabilities is not too significant. 
\begin{figure}
    \centering
    \includegraphics[height=15em,width=22em]{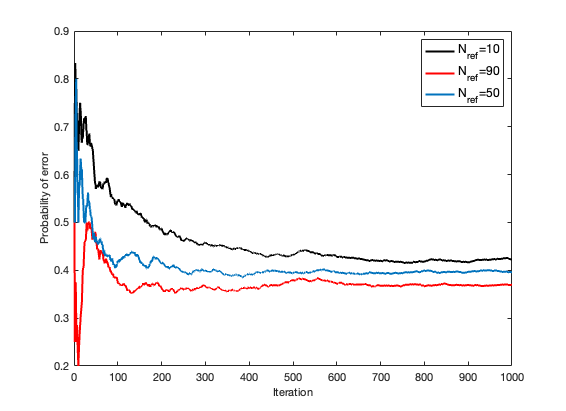}
    \caption{$Pe$ vesus the number of iterations when different values of $N_{ref}$ are utilized for the GLRTRS detector.}
    \label{fig:fig5}
\end{figure}

In Fig.~\ref{fig:fig5}, the error probability and the convergence rate of the GLRTRS detector with a different number of reference nodes are presented. The numbers of sensors used for detection purposes in the GLRTRS detectors with different values of $N_{ref}$ are equal to 200, i.e., $N-N_{ref}=200$. It can be observed that the convergence rate is faster, and the error probability is lower when more reference nodes are used.

\begin{figure}
    \centering
    \includegraphics[height=15em,width=22em]{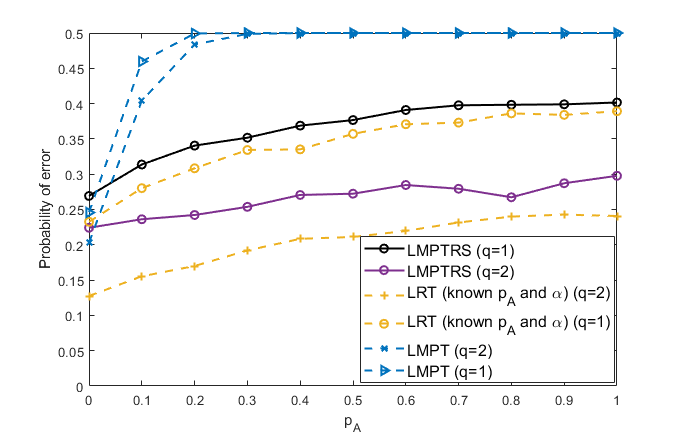}
    \caption{
    Comparison of $Pe$ for the LMPTRS, LRT and quantized LMPT detectors.}
    \label{fig:fig6}
\end{figure}
\begin{figure}
    \centering
    \includegraphics[height=15em,width=22em]{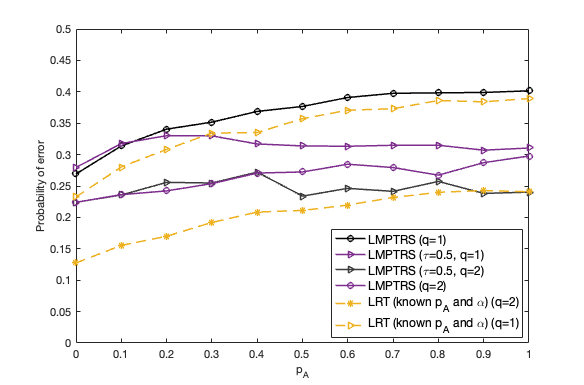}
    \caption{
    $Pe$ vesus $P_A$ when different values of $q$ are utilized for the LMPTRS and the E-LMPTRS detectors.}
    \label{fig:fig7}
\end{figure}

Fig.~\ref{fig:fig6} shows the error probabilities of the LRT detector with perfect knowledge of $\{P_A, \alpha, p\}$, the quantized LMPT detector (proposed in \cite{wang2019distributed}) and the proposed LMPTRS detector for $q=1$ and $q=2$, respectively. We can observe that the quantized LMPT detector proposed in \cite{wang2019distributed} is also extremely vulnerable to attacks for both one-bit and multilevel quantization when all the $p$, $P_A$ and $\alpha$ are unknown. However, it can be observed that when $q=1$, the proposed LMPTRS detector is capable of obtaining an error probability close to the benchmark error probability that is obtained by employing the LRT detector with perfect knowledge of the attack parameters $\{P_A, \alpha, p\}$. Similar to the conclusion we obtained from Fig. \ref{fig:fig3}, the LMPTRS detector outperforms the quantized LMPT detector proposed in \cite{wang2019distributed} in the presence of attacks. The error probability of the proposed LMPTRS detector decreases with increasing $q$, and a higher value of $q$ increases the difference between the benchmark error probability and the proposed LMPTRS detector error probability. 
It is also possible to further reduce the error probability of the system by assuming that the fraction of Byzantine nodes $\alpha$ is known to the system. As shown in Fig.~\ref{fig:fig7}, the E-LMPTRS detector outperforms both the quantized LMPT detector and the benchmark LRT detector with perfect knowledge of the attack parameters by filtering potential Byzantine nodes when $q=1$. When $q$ increases (e.g., $q = 2$), the E-LMPTRS detector still outperforms the quantized LMPT detector.

\section{Conclusion}\label{sec:conclusion}
The distributed detection problem of sparse stochastic signals with quantized measurements in the presence of Byzantine attacks was investigated in this paper. The sparse stochastic signals are characterized by sparsity degree, and the BG distribution was utilized for sparsity modeling. We proposed the LMPTRS and GLRTRS detectors with adaptive thresholds, given that the sparsity degree $p$ and the attack parameters, i.e., $\alpha$ and $P_A$ are unknown. The simulation results showed that the LMPTRS and GLRTRS detectors outperform the LMPT detector under attack and achieve a detection performance close to the benchmark LRT detector with perfect knowledge of the attack parameters and sparsity degree $p$. When the fraction of Byzantines $\alpha$ in the networks is assumed to be known, the E-LMPTRS and E-GLRTRS detectors were proposed to further improve the detection performance of the system by filtering out potential malicious sensors. Simulation results showed that the proposed enhanced detectors outperform LMPTRS and GLRTRS detectors.

In this work, the predefined quantizer thresholds we utilize come from \cite{wang2018detection}. In the future, we intend to consider the optimization of the predefined quantizer thresholds for our proposed detectors.

\appendices
\section{Proof of Theorem}\label{Proof_theo}
We first consider the scenario where sensors send binary decisions to the FC, i.e., $q=1$. After that, we consider the system where sensors send q-bits decisions to the FC ($q\geq 2$). Here, we only consider the assumption that $\widetilde\tau_{j,2^q}\ll\tau_{i,1}$. Nevertheless, we can reach similar conclusions if we assume $\tau_{i,2^q}\ll\widetilde\tau_{j,1}$.

\subsubsection{When sensors send binary decisions (q=1)}

The joint pmf of local decisions coming from reference sensors under hypothesis $\mathcal{H}_h$ is given as
\begin{align}\label{eq:binary_joint}
    P(\mathbf{U}_{ref}|\mathcal{H}_h,p,x)=&\prod_{i=1}^{N_{ref}}(1-x)^{\mathbf{u_i}}x^{1-\mathbf{u_i}}
\end{align}
for $h=0,1$. Take the logarithm of both sides of \eqref{eq:binary_joint}, we have
\begin{align}\label{eq:log_binary_joint}
    \log P(\mathbf{U}_{ref}|\mathcal{H}_h,p,x)=&\sum_{i=1}^{N_{ref}}\left[\mathbf{u_i}\log(1-x)+(1-\mathbf{u_i})\log x\right]\notag\\
    =&Y\log(1-x)+(N_{ref}-Y)\log x,
\end{align}
where $Y=\sum_{i=1}^{N_{ref}}\mathbf{u_i}$. Let $\frac{\partial P(\mathbf{U}_{ref}|\mathcal{H}_h,p,x)}{\partial x}=0$, we are able to obtain the estimated attack parameter $\hat{x}_h$ under hypothesis $\mathcal{H}_h$ which maximizes $\log P(\mathbf{U}_{ref}|\mathcal{H}_h,p,x)$ and the estimated attack parameter $\hat{x}_h$ is given as
\begin{equation}
    \hat{x}_h=1-\frac{Y}{N_{ref}}
\end{equation}

In order to evaluate the estimator performance, it should be noted that it is unbiased since
\begin{align}
    E[\hat{x}_h]=&1-\frac{1}{N_{ref}}E[Y]\notag\\
    =&1-\frac{1}{N_{ref}}\sum_{i=1}^{N_{ref}}E[\mathbf{u_i}]\notag\\
    =&x
\end{align}

The variance of the estimator is given as
\begin{align}\label{eq:variance_estimate1}
    E[\hat{x}_h]=&E[\hat{x}_h^2]-E^2[\hat{x}_h]\notag\\
    =&E\left[\left(1-\frac{Y}{N_{ref}}\right)^2\right]-x^2\notag\\
    =&1-x^2-\frac{2}{N_{ref}}E[Y]+\frac{1}{N_{ref}^2}E[Y^2]\notag\\
    =&1-x^2-2(1-x)+\frac{1}{N_{ref}^2}(Var[Y]+E^2[Y])\notag\\
    =&1-x^2-2(1-x)\notag\\
    &+\frac{1}{N_{ref}^2}[N_{ref}x(1-x)+N_{ref}^2(1-x)^2]\notag\\
    =&\frac{(1-x)x}{N_{ref}}
\end{align}
To evaluate the performance of the estimator, the CRLB can be calculated which is $-\frac{1}{E\left[\partial^2 P(\mathbf{U}_{ref}|\mathcal{H}_h,p,x)/\partial x^2\right]}$. Taking the second derivative of $ P(\mathbf{U}_{ref}|\mathcal{H}_h,p,x)$ with respect to $x$, we have
\begin{align}
    \frac{\partial^2 P(\mathbf{U}_{ref}|\mathcal{H}_h,p,x)}{\partial x^2}=\sum_{i=1}^{N_{ref}}\left[-\frac{\mathbf{u_i}}{(1-x)^2}-\frac{1-\mathbf{u_i}}{x^2}\right].
\end{align}
Subsequently, taking the expectation of the above equation, we have
\begin{align}
    E\left[\frac{\partial^2 P(\mathbf{U}_{ref}|\mathcal{H}_h,p,x)}{\partial x^2}\right]=&\sum_{i=1}^{N_{ref}}E\left[\frac{\partial^2 P(\mathbf{u_i}|\mathcal{H}_h,p,x)}{\partial x^2}\right]\notag\\
    =&\sum_{i=1}^{N_{ref}}-\frac{1}{(1-x)^2}(1-x)-\frac{1}{x^2}x\notag\\
    =&-\frac{N_{ref}}{(1-x)x}.
\end{align}
Therefore, the CRLB is $\frac{(1-x)x}{N_{ref}}$
which is the same as \eqref{eq:variance_estimate1}. This indicates that the proposed estimator attains the CRLB; that is, it is an efficient estimator when sensors in the network send binary decisions.

\subsubsection{When sensors send q-bits decisions ($q\geq2$)}

The joint pmf of local decisions coming from reference sensors under hypothesis $\mathcal{H}_h$ is given as
\begin{align}\label{eq:binary_joint2}
    P(\mathbf{U}_{ref}|\mathcal{H}_h,p,x)=&\prod_{i=1}^{N_{ref}}(1-x)^{I(\mathbf{u_i}=2^q)}\prod_{i=1}^{2^q-1} (\frac{x}{2^q-1})^{I(\mathbf{u_i}=\mathbf{v_j})}
\end{align}
for $h=0,1$. Take the logarithm of both sides of \eqref{eq:binary_joint2}, we have
\begin{align}\label{eq:log_binary_joint2}
    &\log P(\mathbf{U}_{ref}|\mathcal{H}_h,p,x)\notag\\
    =&\sum_{i=1}^{N_{ref}}I(\mathbf{u_i}=2^q)\log(1-x)+\sum_{i=1}^{2^q-1} I(\mathbf{u_i}=\mathbf{v_j})\log(\frac{x}{2^q-1}),
\end{align}

Taking the first derivative of $ P(\mathbf{U}_{ref}|\mathcal{H}_h,p,x)$ with respect to $p$, we have
\begin{align}
    \frac{\partial P(\mathbf{U}_{ref}|\mathcal{H}_h,p,x)}{\partial x}=&\sum_{i=1}^{N_{ref}}\frac{-1}{1-x}I(\mathbf{u_i}=2^q)+\sum_{i=1}^{2^q-1}\frac{1}{x}I(\mathbf{u_i}=\mathbf{v_j})\notag\\
    =&\frac{-Y_1}{1-x}+\frac{Y_2}{x}\label{eq:first_deri_Y_1_Y_2}\\
    =&\frac{-Y_1}{1-x}+\frac{N_{ref}-Y_1}{x}\label{eq:first_deri_Y_1}
\end{align}
where $Y_1=\sum_{i=1}^{N_{ref}}I(\mathbf{u_i}=v_{2^q})$ and $Y_2=\sum_{i=1}^{N_{ref}}\sum_{i=1}^{2^q-1}I(\mathbf{u_i}=\mathbf{v_i})$. In going from \eqref{eq:first_deri_Y_1_Y_2} to \eqref{eq:first_deri_Y_1}, the fact that $Y_1+Y_2=N_{ref}$ is utilized.  Let $\frac{\partial P(\mathbf{U}_{ref}|\mathcal{H}_h,p,x)}{\partial x}=0$, we are able to obtain the estimated attack parameter $\hat{x}$ which maximizes $\log P(\mathbf{U}_{ref}|\mathcal{H}_h,p,x)$. The estimated attack parameter $\hat{x}_h$ under hypothesis $\mathcal{H}_h$ is given as
\begin{equation}
    \hat{x}_h=1-\frac{Y_1}{N_{ref}}
\end{equation}

In order to evaluate the estimator performance, it should be noted that it is unbiased since
\begin{align}
    E[\hat{x}]=&1-\frac{1}{N_{ref}}E[Y_1]\notag\\
    =&1-\frac{1}{N_{ref}}\sum_{i=1}^{N_ref}E[I(\mathbf{u_i}=2^q)]\notag\\
    =&x
\end{align}

Similarly, the variance of the estimator is given as
\begin{align}\label{eq:variance_estimate}
    E[\hat{x}_h]=&E[\hat{x}_h^2]-E^2[\hat{x}_h]\notag\\
    =&E\left[\left(1-\frac{Y_1}{N_{ref}}\right)^2\right]-x^2\notag\\
    =&1-x^2-\frac{2}{N_{ref}}E[Y_1]+\frac{1}{N_{ref}^2}E[Y_1^2]\notag\\
    =&\frac{(1-x)x}{N_{ref}}
\end{align}
To evaluate the performance of the estimator, the CRLB can be calculated which is $-\frac{1}{E\left[\partial^2 P(\mathbf{U}_{ref}|\mathcal{H}_h,p,x)/\partial x^2\right]}$. Taking the second derivative of $ P(\mathbf{U}_{ref}|\mathcal{H}_h,p,x)$ with respect to $p$, we have
\begin{align}
    \frac{\partial^2 P(\mathbf{U}_{ref}|\mathcal{H}_h,p,x)}{\partial x^2}=&\sum_{i=1}^{N_{ref}}-\frac{I(\mathbf{u_i}=2^q)}{(1-x)^2}-\sum_{i=1}^{2^q-1}\frac{I(\mathbf{u_i}=u_j)}{x^2}\notag\\
    =&\sum_{i=1}^{N_{ref}}-\frac{I(\mathbf{u_i}=2^q)}{(1-x)^2}-\frac{1-I(\mathbf{u_i}=2^q)}{x^2}
\end{align}
Subsequently, taking the expectation of the above equation, we have
\begin{align}
    E\left[\frac{\partial^2 P(\mathbf{U}_{ref}|\mathcal{H}_h,p,x)}{\partial x^2}\right]=&\sum_{i=1}^{N_{ref}}E\left[\frac{\partial^2 P(\mathbf{u_i}|\mathcal{H}_h,p,x)}{\partial x^2}\right]\notag\\
    =&\sum_{i=1}^{N_{ref}}-\frac{1}{(1-x)^2}(1-x)-\frac{1}{x^2}x\notag\\
    =&-\frac{N_{ref}}{(1-x)x}
\end{align}
Therefore, the CRLB is $\frac{(1-x)x}{N_{ref}}$
which is the same as \eqref{eq:variance_estimate}. This indicates that the proposed estimator attains the CRLB; that is, it is an efficient estimator when sensors in the network send q-bits decisions. This completes our proof.

\bibliographystyle{IEEEtran}
\bibliography{refer.bib}
\end{document}